\documentclass[fleqn,10pt]{wlscirep}
\title{An Autonomous Programmable Actuator and Shape Reconfigurable Structures using Bistability and Shape Memory Polymers}

\usepackage{siunitx}
\author[1]{Tian Chen}
\author[1,*]{Kristina Shea}
\affil[1]{Engineering Design and Computing Laboratory, D-MAVT, ETH Zurich, 8092, Switzerland}

\affil[*]{kshea@ethz.ch}

\keywords{shape reconfigurable structures, programmable matter, bistable actuator, shape memory polymer, 4D printing}

\begin{abstract}
Autonomous deployment and shape reconfiguration of structures is a crucial field of research in space exploration with emerging applications in the automotive, building and biomedical industries. Challenges in achieving autonomy include: bulky energy sources, imprecise deployment, jamming of components and lack of structural integrity. Leveraging advances in the fields of shape memory polymers, bistability and 3D multi-material printing, we present a 3D printed programmable actuator that enables the autonomous deployment and shape reconfiguration of structures activated though surrounding temperature change. Using a shape memory polymer as the temperature controllable energy source and a bistable mechanism as the linear actuator and force amplifier, the structures achieve precise geometric activation and quantifiable load bearing capacity. The proposed unit actuator integrates these two components and is designed to be assembled into larger deployable and shape reconfigurable structures. First, we demonstrate that the activation of the unit actuator can be sequenced by tailoring each shape memory polymer to a different activation time. Next, by changing the configuration of the actuator, we demonstrate an initially flat surface that transforms into a pyramid or a hyperbolic paraboloid, thus demonstrating a multi-state structure. Load bearing capability is demonstrated for both during activation and in the operating state.
\end{abstract}
\begin{document}

\flushbottom
\maketitle

\thispagestyle{empty}

\begin{figure*}[!ht]
 \centering
 \includegraphics[width=0.475\textwidth]{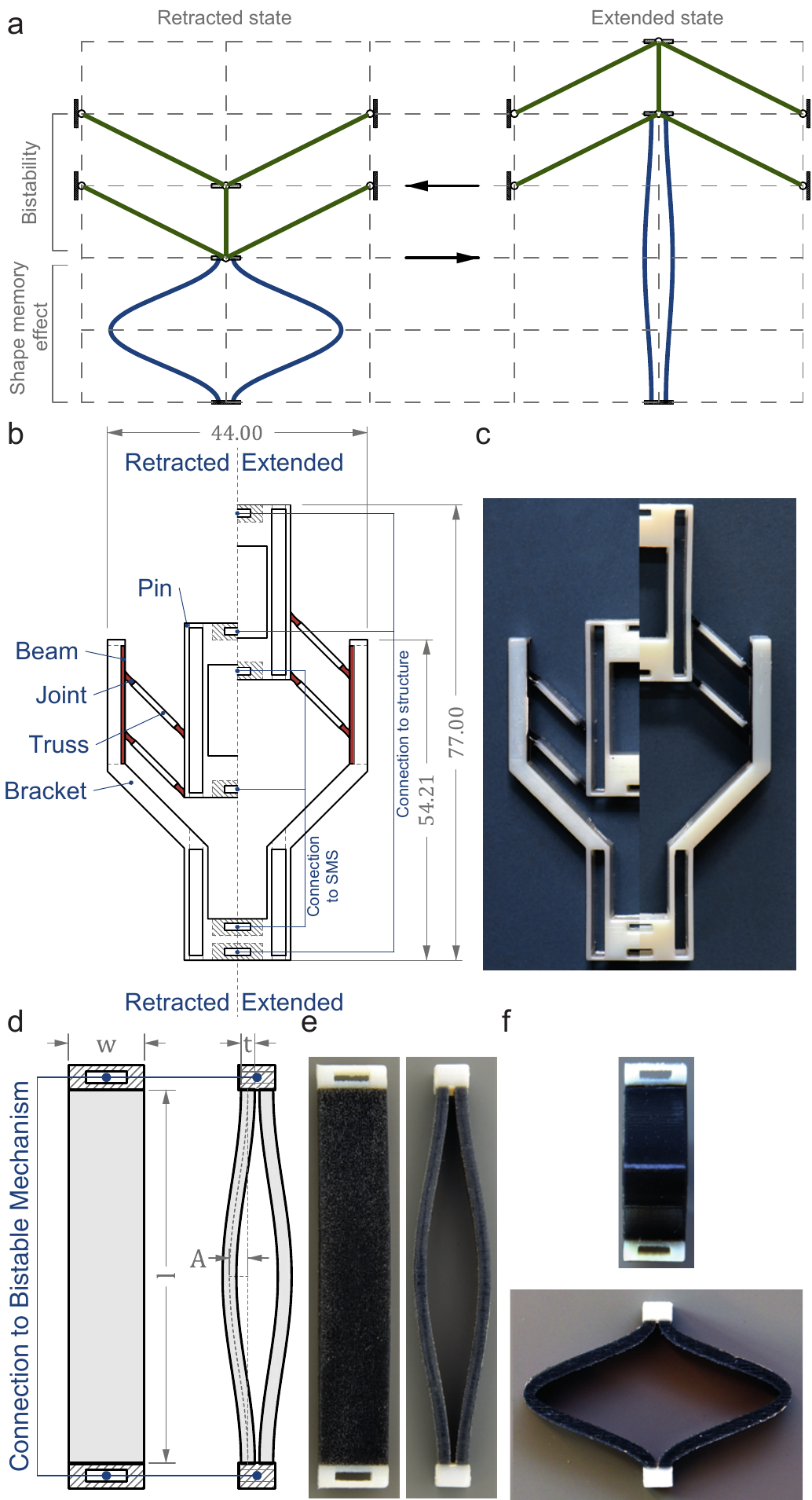}
 \caption{The design of a heat activated bistable unit actuator. a) Schematic showing activation of the actuator. The actuator consists of a bistable mechanism and a shape memory strip (SMS). The SMS provides the force to activate the bistable mechanism. b) The design of the unit actuator shows the bistable mechanism consisting of a bracket that provides structural rigidity, joints that provide the rotational DOF to the bistable trusses, and flexible beams that simulate the boundary condition of the Von Mises Truss. The overall dimensions and the connection points are indicated. d) The SMS whose geometry is parametrically defined to provide both expansion and contraction to the actuator. c,e,f) shows the fabricated specimens.}
 \label{fig:1}
\end{figure*}

\section*{Introduction}
Deployable structures are used in a wide range of applications in space exploration,\cite{Lan2009} biomedical,\cite{Leong2009} and solar energy.\cite{Guo2009} In these applications, the properties of the deployable structures typically sought include large-scale shape change, predictable geometric transformations and reconfigurations, passively controllable activation, load bearing capability and tunable deployment conditions. In addition to motorized actuators, a number of passively actuated designs have been proposed, including electricity,\cite{Wang2014} light,\cite{Ryu2012} pressure,\cite{Maccurdy2016} shape memory effect (SME),\cite{Ge2013} swelling,\cite{Tibbits2014} and piezoelectricity.\cite{Song2007}

In the field of space exploration, deployment is a critical phase in the life cycle of a space structure.\cite{Zolesi2012} With respect to kinematics, entanglement and jamming may occur if part movements cannot be precisely controlled.\cite{Gardsback2007} Mechanically, the driving force must be tuned to overcome resistance at all stages of deployment.\cite{Mitsugi2000a} Traditionally, designs often involve complex packing strategies, such as origami folding,\cite{Miura1985} tensegrity,\cite{Liu2017} as well as bulky powered driving mechanisms such as air pressure\cite{Clem2000} and electrical drives.\cite{Edwards2000a}With multi-material 3D printing technologies, there is potential to fabricate monolithic designs that can be autonomously activated by environmental triggers, e.g. temperature.

An increasing number of active structures are designed and fabricated in a newly coined field of 4D printing.\cite{Tibbits2014} 4D printing uses properties of 3D printing materials to achieve design shape change under environmental forces. The development has been focused in the areas of material synthesis and geometric reconfiguration. In Raviv et al.,\cite{Raviv2014} a hydrogel that swells under water is used as an actuator that changes the shape of a compliant string or surface. Gladman et al.\cite{Gladman2016} continued this method of actuation by designing anisotropic properties of a swelling meta-material. Ge et al.\cite{Ge2016} first utilized the shape memory properties of the photo polymeric inks used in the Polyjet 3D printing technology to achieve shape change.

Each of these actuation methods have areas for improvement. Swelling hydrogel has a long activation time, imprecise geometric configuration change and lack of load bearing capacity. Additionally, it is often impractical to submerge a structure under water. Shape memory polymers, as implemented by Ge et al.,\cite{Ge2016} require the design to be fabricated in the activated state, while offering no real precision in the programmed state.

SME of polymers is adopted in this research as it is able to produce a large actuation strain,\cite{Ding2017} moderate recovery stress, and is able to be fabricated with a 3D printer. By combining this polymer with a bistable mechanism called Von Mises Truss, we are able to define two distinct equilibrium states that can be precisely achieved. Von Mises Type bistable mechanisms have previously been used in active masts,\cite{Schioler2007} for energy absorption,\cite{Shan2015} and more recently to create 3D printed structures.\cite{Chen2017}

In the activation phase, the bistable mechanism acts as a force amplifier, and renders the deployable structure load bearing. By combining the mechanics of bistability and shape memory polymers, we present a unit actuator that delivers the above characteristics and addresses the shortcomings. We propose to integrate this actuator in deployable structures as the source of activation for shape reconfiguration. The structures can be deployed from 2D to 3D states under a certain temperature. By demonstrating a synclastic and an anticlastic structure reconfigured from the same base design, we demonstrate both autonomous deployment and shape reconfiguration. 

\begin{figure*}[!ht]
 \centering
 \includegraphics[width=0.55\textwidth]{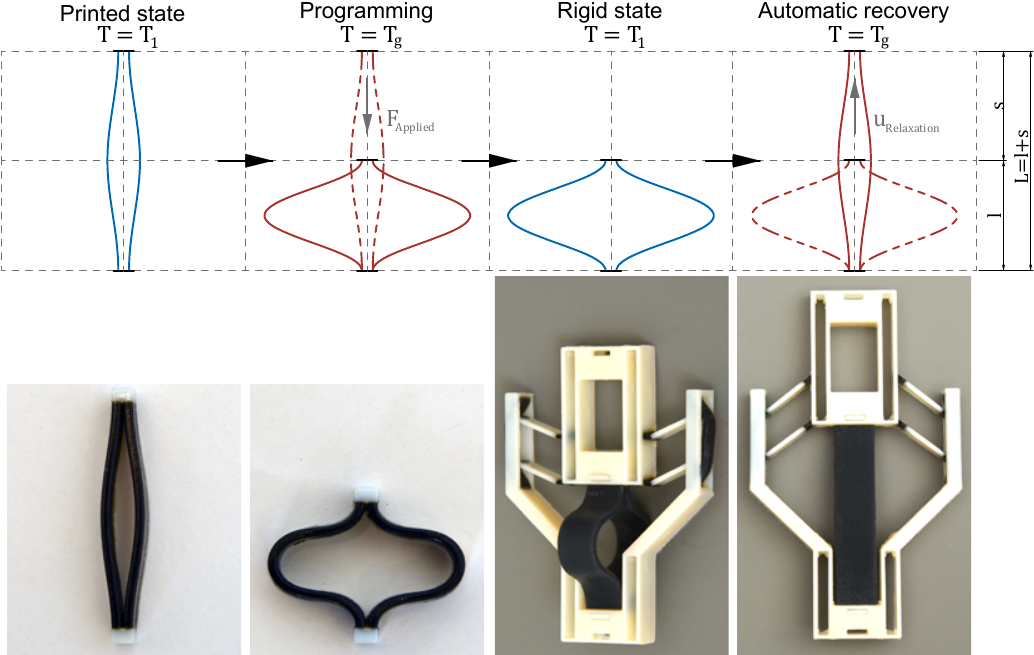}
 \caption{The kinematics and activation sequence of the unit actuator is shown with the expanded actuator as an example. a) The printed state of an expanded SMS. b) By heating the SMS past its $T_\mathrm{g}$ and applying a mechanical force, the SMS is deformed by the stroke length $s$ to its programmed state. c) The programmed SMS is cooled down and assembled into the bistable mechanism. d) The unit actuator is activated through heating. Upon cooling, the final activated state is rigid and behaves as a static member.}
 \label{fig:2}
\end{figure*}

\section*{Results}
\subsection*{Design of a programmable unit actuator using bistability and shape memory polymer}
First, we introduce the geometry of the actuator design, second, we discuss the kinematics and activation sequence, and lastly we investigate its mechanical behavior. For actuation, we propose a unit actuator consisting of a bistable mechanism that dictates the equilibrium states and shape change and the shape memory strip (SMS) that provides the activation force. This unit actuator is capable of activating independently of the rest of the structure. The principle behind this proposed actuator is shown in Figure~\ref{fig:1}a, where the shape memory effect is used to trigger bistability from the contracted to the expanded configuration or vice versa. The bistable mechanism is a realization of the Von Mises Truss~\cite{Mises1923} and is adapted from designs proposed by Chen et al.\cite{Chen2017}

The bistable mechanism is fabricated using both a rigid and a compliant material. Bistability is provided by the rigid truss members, the flexible joint and the flexible beam, which simulates the ideal boundary condition (Figure~\ref{fig:1}b). To achieve predictable deployment, the constituting printed materials are studied using Differential Scanning Calorimetry (DSC). It is found that the glass transition temperature $T_\mathrm{g}$ of the rigid and the compliant components are approximately \SI{80}{\degreeCelsius} and \SI{-5}{\degreeCelsius} respectively.\cite{Wagner2017} This effectively renders both thermally stable at a temperature in the vicinity of the $T_\mathrm{g}$ of the SMS material, which is between \num{30} and \SI{40}{\degreeCelsius}.

The design is fabricated in either the contracted or the expanded configuration and is triggered by the SMS installed beneath the pin. The SMS (Figure~\ref{fig:1}d) consists of two thin, symmetrical strips with a rectangular cross section. The force it delivers is parametrized by its amplitude $A$ and thickness $t$. The stroke length of the bistable mechanism dictates the change in length of the SMS between the fabricated and the activated states. Figure~\ref{fig:1}c shows an example of the unit actuator with the bistable mechanism. While this design can be fabricated in one piece, we choose to separate the bistable mechanism from the SMS so that we are able to investigate each separately.

\begin{figure*}[!ht]
 \centering
 \includegraphics[width=0.95\textwidth]{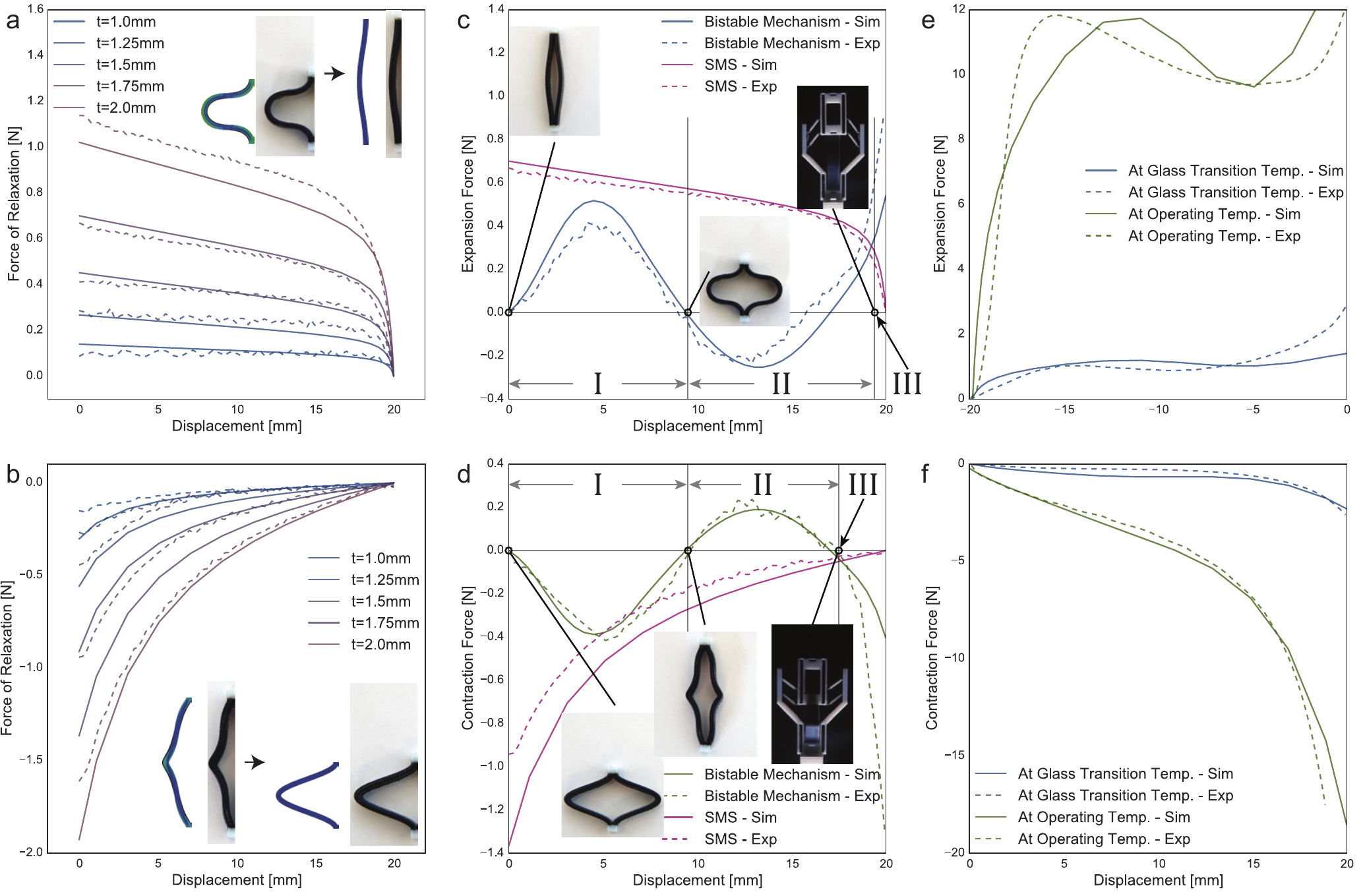}
 \caption{Mechanical characterization of the unit actuator, both in expansion (a, c, e) and contraction (b,d,f), including simulation and experimental results in both the glassy and the rubbery regime. a,b) Parametric study of the influence of the SMS thickness and recovery direction on its activation force at a temperature load above $T_\mathrm{g}$. The range of the thickness studies is between \num{1.0} to \SI{2.0}{\milli\metre} in \SI{0.25}{\milli\metre} intervals. c,d) Shows the force displacement curves of the bistable mechanism and the chosen SMS. Activation is guaranteed when the SMS force is greater than the bistable force in region I. In region II, both SMS and bistability are acting in the same direction. Stable equilibrium is indicated at point III. e,f) These plots compare the overall behavior of the actuator at $T_\mathrm{g}$ and at operating temperature of approximately \SI{20}{\degreeCelsius}.}
 \label{fig:3}
\end{figure*}

The deployment of the unit actuator procedure consists of two phases, the first phase is the programming and assembly of the SMS with the bistable mechanism. The second is the constrained recovery of the SMS and consequently the activation of the unit actuator. In the programming phase, the SMS is heated past $T_\mathrm{g}$ and is either stretched or compressed by a distance equal to the stroke length of the bistable mechanism (Figure~\ref{fig:2}a to b). While confined, the SMS is cooled and installed in the bistable mechanism (Figure~\ref{fig:2}c). The second phase is triggered by raising the temperature of the unit actuator past $T_\mathrm{g}$ (Figure~\ref{fig:2}d). As the SMS recovers, it triggers the bistable mechanism and achieves the deployed state. Once cooled, this activated state behaves as a rigid structure as the SMS returns to its glassy state.

To ensure deployment, the SMS must overcome the activation force of the bistable mechanism at all stages of the deployment process. Mechanical testing is done using a dynamic testing machine with an embedded heat chamber. To simulate the test condition, a Finite Element simulation is performed to obtain the behavior of SMS using a linear viscoelastic constitutive model constructed for the shape memory material. Stress and strain behavior is obtained from Chen et al.\cite{Chen2017} Simulation of the bistable mechanism follows a static non-linear FEA with prescribed displacement using beam elements.

Simulation and experimental results are shown in Figure~\ref{fig:3}. Panels a and b show the force exerted by the SMS during relaxation when $T\ge T_\mathrm{g}$. By varying the thickness of the SMS, we are able to tune the force from \num{0.1} to \SI{1}{\newton} during expansion and \num{0.1} to \SI{2}{\newton} for contraction. Using this data, we select a SMS with the force required to activate the bistable mechanism. As shown in Panels c and d, to activate the bistable mechanism when $T\ge T_\mathrm{g}$, the SMS must deliver a constrained relaxation force that is greater than the bistability activation force, i.e. $F_\mathrm{SMS}-F_\mathrm{bi} \ge 0$. This must be explicitly satisfied in Region I where the bistable force acts in opposition to the SMS. In Region II, after the bistable mechanism triggers, both forces act in the same direction. This enables the force amplification characteristic of the bistable mechanism. Equilibrium occurs at the end of Region II where the two forces balance once again (labelled as III). Note that this is slightly before full stroke length as the bistable mechanism has an asymmetrical behavior.

Panels e and f show the overall mechanical behavior of a unit actuator under either activation temperature (i.e. $T \ge T_\mathrm{g}$) or operating temperature (i.e. $T<T_\mathrm{g}$, in this case, $T \sim \SI{22}{\degreeCelsius}$). In the first instance, the overall behavior equals to the sum of the relaxation force of the SMS and the activation force of the bistable mechanism. In the second instance, by cooling the SMS to its glassy state, its Young's modulus increases significantly. As a result, the unit actuator behaves as if it were a static structural element, allowing a much stiffer force displacement response.

\begin{figure*}[!ht]
 \centering
 \includegraphics[width=0.55\textwidth]{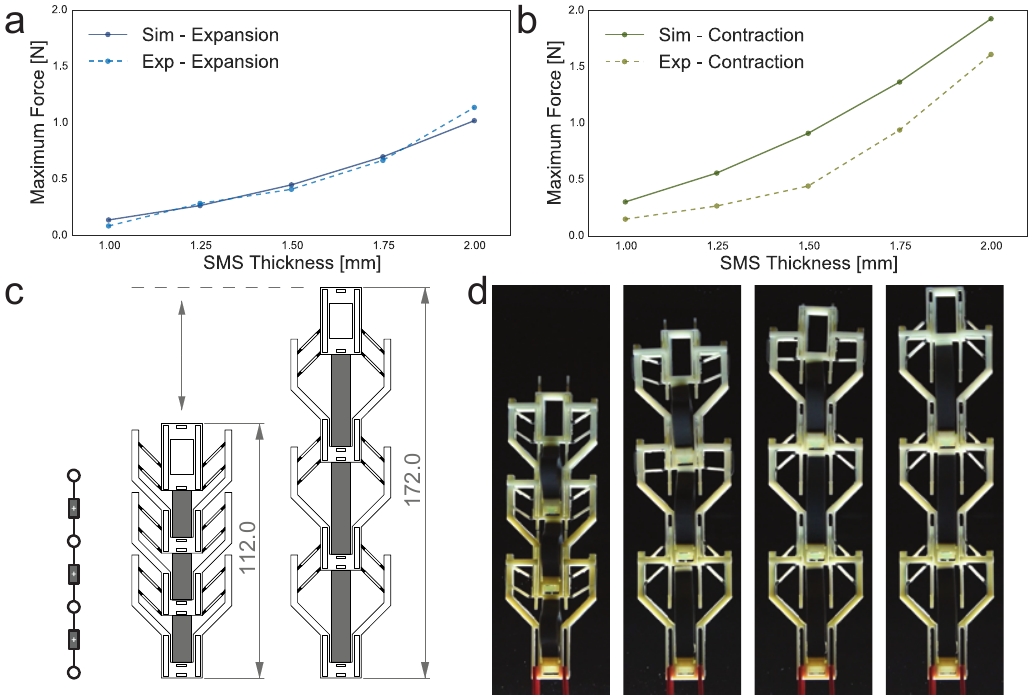}
 \caption{Sequenced activation of multiple actuators. a,b) Plots showing the maximum activation force of the different SMS in expansion and contraction direction respectively. c) Schematic showing three expansion actuators connected in series (+) and the physical specimen in both initial and final state. d) Video screen captures showing the sequenced activation of the actuators.}
 \label{fig:4}
\end{figure*}

We demonstrate that the actuator can be integrated into a larger structure by showing the controlled activation of three serially connected actuators. By varying the thickness of the SMS, we can control the exerted force (Figure~\ref{fig:4}a,b) and thus the sequence of activation due to heat conduction of the shape memory polymer. SMS with thicknesses of \num{1.50}, \num{1.75}, \SI{2.0}{\milli\metre} are used as they can all trigger the bistable element, yet have different activation times. The actuator is designed such that its pin is part of the frame of a second, serially connected actuator (Figure~\ref{fig:4}c). This increases the overall expansion ratio as more actuators are linked. The activation sequence begins with the thinnest SMS and completes with the thickest SMS (Figure~\ref{fig:4}d).

\subsection*{Design of Deployable Structures}
\begin{figure*}[!ht]
 \centering
 \includegraphics[width=0.95\textwidth]{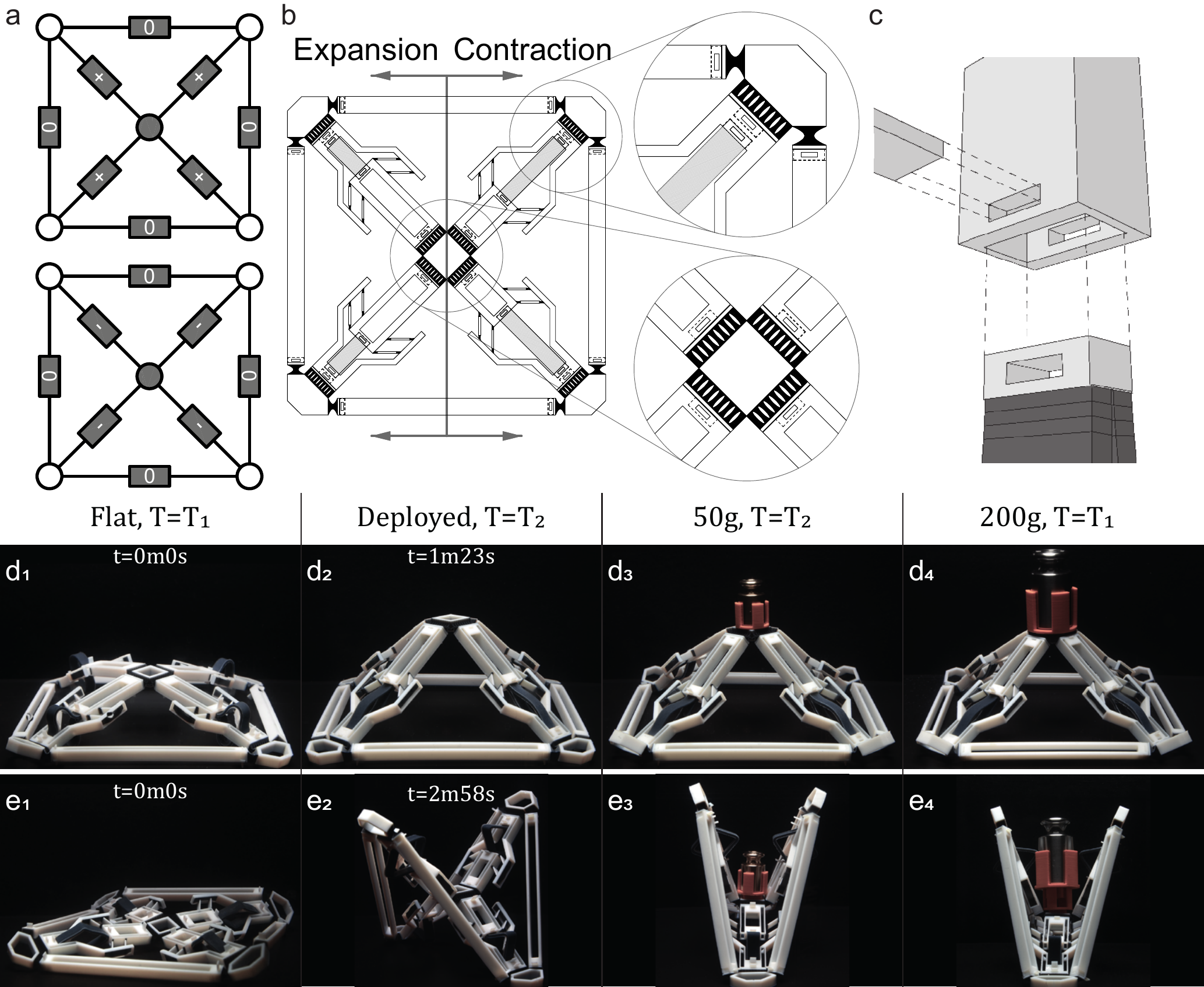}
 \caption{Shape reconfigurable structures showing activation and load bearing capability. a) Schematic of the designs. Four actuators are integrated in a flexible frame, the first has four expansion actuators (+), and the second has four contraction actuators (-). b) The physical design of the structures, the compliant joints are fabricated with a flexible material, and functionally graded to provide the desired flexibility and robustness. c) A detailed view of a universal connector between all parts of the assembly. d) and e) show video screen captures of the expanding and contracting structural deployment respectively, 1. Flat state, 2. Activated state. 3. Load bearing capacity under high temperature, 4. Load bearing capacity under operating temperature.}
 \label{fig:5}
\end{figure*}

We now integrate the described actuators in 2D designs that deploy into load carrying 3D structures. An edge-node schematic of the structure is shown in Figure~\ref{fig:5}a. The design can be simulated as a truss network, i.e. the edges can only sustain axial load and the nodes are pin jointed. The design consists of a rectangular frame with four cross edges. Each edge can assume three behaviors: expansion, contraction or remain static. In this work, two physical realizations of the design are demonstrated by switching the initial configuration of the cross edges between expansion and contraction actuators (Figure~\ref{fig:5}b). Compliant joints are placed at the nodes where rotational Degrees of Freedom (DOF) are required. The stiffness of the joints is reduced by fabricating them with a soft elastomer and internal microstructure. The different parts are assembled with a common joint (Figure~\ref{fig:5}c).

Each SMS is programmed separately and assembled into the structure. The initial state is planar. For deployment, the structure is submerged under \SI{40}{\degreeCelsius} water. Self-weight of the structures is neglected as the density of the materials are approximately \SI{1175}{\kilo\gram\per\cubic\meter}, which is slightly higher than water. Both structures activate as predicted. With the expanding structure, a synclastic surface is formed resembling a pyramid. With the contracting structure, an anticlastic surface is formed. The timing of the activation for both is under three minutes, which is significantly faster than using swelling.

In addition to activation, two load cases are tested by placing a specified mass at the apex of each structure. The first is under the deployment condition where the temperature remains at \SI{40}{\degreeCelsius}. The load sustained is \SI{0.050}{\kilogram} or an equivalent weight of \SI{0.430}{\newton}. The second load case is conducted after the surrounding temperature lowers to approximately \SI{20}{\degreeCelsius}. The load sustained is \SI{0.200}{\kilogram} or an equivalent weight of \SI{1.70}{\newton}.

\section*{Discussion}
The presented work demonstrates the design of autonomous shape reconfigurable structures. We propose a programmable unit actuator that either expands or contracts under temperature change. Using the viscoelastic behavior of thermosetting polymers, a pre-strain is imposed on the Shape Memory Strips (SMS) at a temperature greater than the polymer's glass transition temperature, $T_\mathrm{g}$. This pre-strain locks in when the temperature lowers and is released when the temperature increases again, thereby triggering the bistable mechanism. Using this actuator, we demonstrate the deployment of 2D sequenced actuators and load bearing 3D space structures.

The proposed unit actuator can be integrated in structural applications where the transport volume must be much smaller than the operating volume. With respect to the example of deployable space structures, the proposed design has a number of advantages compared to the state-of-art in all phases of the deployment process.

In the fabrication phase, both the bistable mechanism and the SMS are 3D printed with minimal support material and assembled with a universal connector (Figure~\ref{fig:5}c). The activation force of the bistable mechanism is dependent on its joint length and stiffness.\cite{Chen2017} The triggering temperature and duration of the SMS can be tuned by varying the chemical composition of the shape memory polymer resin~\cite{Ge2016} and by changing its thickness respectively.

In transportation and the deployment phase, the stowed state of the proposed actuator is precisely defined by the first equilibrium point of the bistable mechanism. Designs that feature only SMPs are programmed using ad-hoc methods such as hand pulling.\cite{Ge2014a} This flat stowed state is maintained without a restraining mechanism.\cite{Brinkmeyer2015} Rather than motorized deployment, the proposed structures are activated through environmental change, thereby removing the need for electrical energy input and eliminating the possibility of jamming of parts.

In the operating phase, the structure is load bearing when deployed, as the bistable mechanism amplifies the load bearing capacity of the unit actuator independently of the surrounding temperature. This is in contrast with most 4D printed designs that focus solely on shape transformation rather than functionality in the final state.\cite{Tibbits2014} Under operating temperature (i.e. $T<T_\mathrm{g}$), the SMS returns to its glassy state and behaves as a rigid member, effectively making the unit actuator a static structural element.

As demonstrated, the structures can reconfigure themselves both positively and negatively, with a combination of these two modes and given many actuators, one can form a surface of almost any given curvature. As we have characterized the thermal mechanical behavior of the unit actuator, simulation of the deployment and load capacity can be done using form-finding algorithms, developed in previous work, that can be extended to generate and optimize large scale reconfigurable structures.\cite{Chen2017} 

\section*{Methods}
All designs in this work are fabricated with a Stratasys Objet3 Connex500 multi-material printer. The three materials used to fabricate the presented designs are FLX9895, the shape memory polymer, RGD525, a temperature resistant rigid plastic and Agilus30, an elastomer-like material.\cite{stratasys2014} The shape memory polymer is characterized using Differential Scanning Calorimetry. The unit actuators are characterized using the Instron E3000 Dynamic testing machine with a thermal chamber. Simulations are performed using Abaqus 14.1. Testing of the deployable structures are done under water that is heated with a heating element.


\begin{thebibliography}{10}
\expandafter\ifx\csname url\endcsname\relax
  \def\url#1{\texttt{#1}}\fi
\expandafter\ifx\csname urlprefix\endcsname\relax\def\urlprefix{URL }\fi
\expandafter\ifx\csname doiprefix\endcsname\relax\def\doiprefix{DOI }\fi
\providecommand{\bibinfo}[2]{#2}
\providecommand{\eprint}[2][]{\url{#2}}

\bibitem{Lan2009}
\bibinfo{author}{Lan, X.} \emph{et~al.}
\newblock \bibinfo{journal}{\bibinfo{title}{{Fiber reinforced shape-memory
  polymer composite and its application in a deployable hinge}}}.
\newblock {\emph{\JournalTitle{Smart Materials and Structures}}}
  \textbf{\bibinfo{volume}{18}}, \bibinfo{pages}{024002}
  (\bibinfo{year}{2009}).
\newblock \doiprefix 10.1088/0964-1726/18/2/024002.

\bibitem{Leong2009}
\bibinfo{author}{Leong, T.~G.} \emph{et~al.}
\newblock \bibinfo{journal}{\bibinfo{title}{{Tetherless thermobiochemically
  actuated microgrippers.}}}
\newblock {\emph{\JournalTitle{Proceedings of the National Academy of Sciences
  of the United States of America}}} \textbf{\bibinfo{volume}{106}},
  \bibinfo{pages}{703--708} (\bibinfo{year}{2009}).
\newblock \doiprefix 10.1073/pnas.0807698106.

\bibitem{Guo2009}
\bibinfo{author}{Guo, X.} \emph{et~al.}
\newblock \bibinfo{journal}{\bibinfo{title}{{Two- and three-dimensional folding
  of thin film single-crystalline silicon for photovoltaic power
  applications.}}}
\newblock {\emph{\JournalTitle{Proceedings of the National Academy of Sciences
  of the United States of America}}} \textbf{\bibinfo{volume}{106}},
  \bibinfo{pages}{20149--54} (\bibinfo{year}{2009}).
\newblock \doiprefix 10.1073/pnas.0907390106.

\bibitem{Wang2014}
\bibinfo{author}{Wang, Q.}, \bibinfo{author}{Gossweiler, G.~R.},
  \bibinfo{author}{Craig, S.~L.} \& \bibinfo{author}{Zhao, X.}
\newblock \bibinfo{journal}{\bibinfo{title}{{Cephalopod-inspired design of
  electro-mechano-chemically responsive elastomers for on-demand fluorescent
  patterning.}}}
\newblock {\emph{\JournalTitle{Nature communications}}}
  \textbf{\bibinfo{volume}{5}}, \bibinfo{pages}{4899} (\bibinfo{year}{2014}).
\newblock \doiprefix 10.1038/ncomms5899.
\newblock \eprint{arXiv:1507.02142v2}.

\bibitem{Ryu2012}
\bibinfo{author}{Ryu, J.} \emph{et~al.}
\newblock \bibinfo{journal}{\bibinfo{title}{{Photo-origami-Bending and folding
  polymers with light}}}.
\newblock {\emph{\JournalTitle{Applied Physics Letters}}}
  \textbf{\bibinfo{volume}{100}} (\bibinfo{year}{2012}).
\newblock \doiprefix 10.1063/1.3700719.

\bibitem{Maccurdy2016}
\bibinfo{author}{Maccurdy, R.}, \bibinfo{author}{Katzschmann, R.},
  \bibinfo{author}{Kim, Y.} \& \bibinfo{author}{Rus, D.}
\newblock \bibinfo{journal}{\bibinfo{title}{{Printable hydraulics: A method for
  fabricating robots by 3D co-printing solids and liquids}}}.
\newblock {\emph{\JournalTitle{Proceedings - IEEE International Conference on
  Robotics and Automation}}} \textbf{\bibinfo{volume}{2016-June}},
  \bibinfo{pages}{3878--3885} (\bibinfo{year}{2016}).
\newblock \doiprefix 10.1109/ICRA.2016.7487576.
\newblock \eprint{1512.03744}.

\bibitem{Ge2013}
\bibinfo{author}{Ge, Q.}, \bibinfo{author}{Qi, H.~J.} \& \bibinfo{author}{Dunn,
  M.~L.}
\newblock \bibinfo{journal}{\bibinfo{title}{{Active materials by four-dimension
  printing}}}.
\newblock {\emph{\JournalTitle{Applied Physics Letters}}}
  \textbf{\bibinfo{volume}{103}} (\bibinfo{year}{2013}).
\newblock \doiprefix 10.1063/1.4819837.

\bibitem{Tibbits2014}
\bibinfo{author}{Tibbits, S.}
\newblock \bibinfo{journal}{\bibinfo{title}{{4D Printing: Multi-Material Shape
  Change}}}.
\newblock {\emph{\JournalTitle{Architectural Design: High Definition: Zero
  Tolerance in Design and Production}}} \textbf{\bibinfo{volume}{84}},
  \bibinfo{pages}{116--121} (\bibinfo{year}{2014}).
\newblock \doiprefix 10.1002/ad.1710.

\bibitem{Song2007}
\bibinfo{author}{Song, Y.~S.} \& \bibinfo{author}{Sitti, M.}
\newblock \bibinfo{journal}{\bibinfo{title}{{Surface-tension-driven
  biologically inspired water strider robots: Theory and experiments}}}.
\newblock {\emph{\JournalTitle{IEEE Transactions on Robotics}}}
  \textbf{\bibinfo{volume}{23}}, \bibinfo{pages}{578--589}
  (\bibinfo{year}{2007}).
\newblock \doiprefix 10.1109/TRO.2007.895075.

\bibitem{Zolesi2012}
\bibinfo{author}{Zolesi, V.} \emph{et~al.}
\newblock \bibinfo{journal}{\bibinfo{title}{{On an innovative deployment
  concept for large space structures}}}.
\newblock {\emph{\JournalTitle{42nd International Conference on Environmental
  Systems}}} \bibinfo{pages}{1--14} (\bibinfo{year}{2012}).
\newblock \doiprefix 10.2514/6.2012-3601.

\bibitem{Gardsback2007}
\bibinfo{author}{G{\"{a}}rdsback, M.}, \bibinfo{author}{Tibert, G.} \&
  \bibinfo{author}{Izzo, D.}
\newblock \bibinfo{journal}{\bibinfo{title}{{Design considerations and
  deployment simulations of spinning space webs}}}.
\newblock {\emph{\JournalTitle{48th AIAA/ASME/ASCE/AHS/ASC Structures,
  Structural Dynamics, and Materials Conference}}}
  \textbf{\bibinfo{volume}{2}}, \bibinfo{pages}{1503--1512}
  (\bibinfo{year}{2007}).
\newblock \doiprefix 10.2514/6.2007-1829.

\bibitem{Mitsugi2000a}
\bibinfo{author}{Mitsugi, J. I.~N.}, \bibinfo{author}{Ando, K.},
  \bibinfo{author}{Senbokuya, Y.} \& \bibinfo{author}{Meguro, A.}
\newblock \bibinfo{journal}{\bibinfo{title}{{Deployment Analysis of Large Space
  Antenna Using Flexible Multibody Dynamics Simulation}}}.
\newblock {\emph{\JournalTitle{Acta Astronautica}}}
  \textbf{\bibinfo{volume}{47}}, \bibinfo{pages}{19--26}
  (\bibinfo{year}{2000}).

\bibitem{Miura1985}
\bibinfo{author}{Miura, K.}
\newblock \bibinfo{journal}{\bibinfo{title}{{Method of packaging and deployment
  of large membranes in space}}}.
\newblock {\emph{\JournalTitle{The Institute of Space and Astronautical Science
  report}}} \textbf{\bibinfo{volume}{618}}, \bibinfo{pages}{1--9}
  (\bibinfo{year}{1985}).
\newblock \doiprefix 10.1177/0096340215581359.

\bibitem{Liu2017}
\bibinfo{author}{Liu, K.}, \bibinfo{author}{Wu, J.}, \bibinfo{author}{Paulino,
  G.~H.} \& \bibinfo{author}{Qi, H.~J.}
\newblock \bibinfo{journal}{\bibinfo{title}{{Programmable Deployment of
  Tensegrity Structures by Stimulus-Responsive Polymers}}}.
\newblock {\emph{\JournalTitle{Scientific Reports}}}
  \textbf{\bibinfo{volume}{7}}, \bibinfo{pages}{3511} (\bibinfo{year}{2017}).
\newblock \doiprefix 10.1038/s41598-017-03412-6.

\bibitem{Clem2000}
\bibinfo{author}{Clem, A.}, \bibinfo{author}{Smith, S.} \&
  \bibinfo{author}{Main, J.}
\newblock \bibinfo{journal}{\bibinfo{title}{{A pressurized deployment model for
  inflatable space structures}}}.
\newblock {\emph{\JournalTitle{41st Structures, Structural Dynamics, and
  Materials Conference and Exhibit}}}  (\bibinfo{year}{2000}).
\newblock \doiprefix 10.2514/6.2000-1808.

\bibitem{Edwards2000a}
\bibinfo{author}{Edwards, B.~C.}
\newblock \bibinfo{journal}{\bibinfo{title}{{Design and deployment of a space
  elevator}}}.
\newblock {\emph{\JournalTitle{Acta Astronautica}}}
  \textbf{\bibinfo{volume}{47}}, \bibinfo{pages}{735--744}
  (\bibinfo{year}{2000}).
\newblock \doiprefix 10.1016/S0094-5765(00)00111-9.

\bibitem{Raviv2014}
\bibinfo{author}{Raviv, D.} \emph{et~al.}
\newblock \bibinfo{journal}{\bibinfo{title}{{Active printed materials for
  complex self-evolving deformations}}}.
\newblock {\emph{\JournalTitle{Scientific Reports}}}
  \textbf{\bibinfo{volume}{4}}, \bibinfo{pages}{7422} (\bibinfo{year}{2014}).
\newblock \doiprefix 10.1038/srep07422.

\bibitem{Gladman2016}
\bibinfo{author}{{Sydney Gladman}, A.}, \bibinfo{author}{Matsumoto, E.~A.},
  \bibinfo{author}{Nuzzo, R.~G.}, \bibinfo{author}{Mahadevan, L.} \&
  \bibinfo{author}{Lewis, J.~A.}
\newblock \bibinfo{journal}{\bibinfo{title}{{Biomimetic 4D printing.}}}
\newblock {\emph{\JournalTitle{Nature materials}}}
  \textbf{\bibinfo{volume}{15}}, \bibinfo{pages}{413--8}
  (\bibinfo{year}{2016}).
\newblock \doiprefix 10.1038/nmat4544.
\newblock \eprint{arXiv:1011.1669v3}.

\bibitem{Ge2016}
\bibinfo{author}{Ge, Q.} \emph{et~al.}
\newblock \bibinfo{journal}{\bibinfo{title}{{Multimaterial 4D Printing with
  Tailorable Shape Memory Polymers}}}.
\newblock {\emph{\JournalTitle{Scientific Reports}}}
  \textbf{\bibinfo{volume}{6}}, \bibinfo{pages}{31110} (\bibinfo{year}{2016}).
\newblock \doiprefix 10.1038/srep31110.

\bibitem{Ding2017}
\bibinfo{author}{Ding, Z.} \emph{et~al.}
\newblock \bibinfo{journal}{\bibinfo{title}{{Direct 4D printing via active
  composite materials}}}.
\newblock {\emph{\JournalTitle{Science Advances}}}
  \textbf{\bibinfo{volume}{3}}, \bibinfo{pages}{e1602890}
  (\bibinfo{year}{2017}).
\newblock \doiprefix 10.1126/sciadv.1602890.

\bibitem{Schioler2007}
\bibinfo{author}{Schioler, T.} \& \bibinfo{author}{Pellegrino, S.}
\newblock \bibinfo{journal}{\bibinfo{title}{{Space Frames with Multiple Stable
  Configurations}}}.
\newblock {\emph{\JournalTitle{AIAA Journal}}} \textbf{\bibinfo{volume}{45}},
  \bibinfo{pages}{1740--1747} (\bibinfo{year}{2007}).
\newblock \doiprefix 10.2514/1.16825.

\bibitem{Shan2015}
\bibinfo{author}{Shan, S.} \emph{et~al.}
\newblock \bibinfo{journal}{\bibinfo{title}{{Multistable Architected Materials
  for Trapping Elastic Strain Energy}}}.
\newblock {\emph{\JournalTitle{Advanced Materials}}}
  \textbf{\bibinfo{volume}{27}}, \bibinfo{pages}{4296--4301}
  (\bibinfo{year}{2015}).
\newblock \doiprefix 10.1002/adma.201501708.

\bibitem{Chen2017}
\bibinfo{author}{Chen, T.}, \bibinfo{author}{Mueller, J.} \&
  \bibinfo{author}{Shea, K.}
\newblock \bibinfo{journal}{\bibinfo{title}{{Integrated Design and Simulation
  of Tunable, Multi-State Structures Fabricated Monolithically with
  Multi-Material 3D Printing}}}.
\newblock {\emph{\JournalTitle{Scientific Reports}}}
  \textbf{\bibinfo{volume}{7}}, \bibinfo{pages}{45671} (\bibinfo{year}{2017}).
\newblock \doiprefix 10.1038/srep45671.

\bibitem{Mises1923}
\bibinfo{author}{von Mises, R.}
\newblock \bibinfo{journal}{\bibinfo{title}{{{\"{U}}ber die
  Stabilit{\"{a}}tsprobleme der Elastizit{\"{a}}tstheorie}}}.
\newblock {\emph{\JournalTitle{Zeitschrift fur angewandte Mathematik und
  Mechanik}}} \textbf{\bibinfo{volume}{3}}, \bibinfo{pages}{406--422}
  (\bibinfo{year}{1923}).

\bibitem{Wagner2017}
\bibinfo{author}{Wagner, M.}, \bibinfo{author}{Chen, T.} \&
  \bibinfo{author}{Shea, K.}
\newblock \bibinfo{journal}{\bibinfo{title}{{Large Shape Transforming 4D
  Auxetic Structures}}}.
\newblock {\emph{\JournalTitle{3D Printing and Additive Manufacturing}}}
  \textbf{\bibinfo{volume}{September}} (\bibinfo{year}{2017}).

\bibitem{Ge2014a}
\bibinfo{author}{Ge, Q.}, \bibinfo{author}{Dunn, C.~K.}, \bibinfo{author}{Qi,
  H.~J.} \& \bibinfo{author}{Dunn, M.~L.}
\newblock \bibinfo{journal}{\bibinfo{title}{{Active origami by 4D printing}}}.
\newblock {\emph{\JournalTitle{Smart Materials and Structures}}}
  \textbf{\bibinfo{volume}{23}}, \bibinfo{pages}{1--15} (\bibinfo{year}{2014}).
\newblock \doiprefix 10.1088/0964-1726/23/9/094007.

\bibitem{Brinkmeyer2015}
\bibinfo{author}{Brinkmeyer, A.}, \bibinfo{author}{Pellegrino, S.} \&
  \bibinfo{author}{Weaver, P.~M.}
\newblock \bibinfo{journal}{\bibinfo{title}{{Effects of Long-Term Stowage on
  the Deployment of Bistable Tape Springs}}}.
\newblock {\emph{\JournalTitle{Journal of Applied Mechanics}}}
  \textbf{\bibinfo{volume}{83}}, \bibinfo{pages}{011008}
  (\bibinfo{year}{2015}).
\newblock \doiprefix 10.1115/1.4031618.

\bibitem{stratasys2014}
\bibinfo{author}{{Stratasys Ltd}}.
\newblock \bibinfo{title}{{Materials data sheet}}.
\newblock \bibinfo{type}{Tech. Rep.} (\bibinfo{year}{2014}).

\end{thebibliography}
\end{document}